\documentclass[12pt]{elsart}
\usepackage[T2A]{fontenc}
\usepackage{amsmath}
\usepackage{amssymb}
\usepackage{graphicx}

\DeclareMathSymbol{\lang}{\mathord}{symbols}{"68}
\DeclareMathSymbol{\rang}{\mathord}{symbols}{"69}
\DeclareMathSymbol{\openbra}{\mathord}{symbols}{"68}
\DeclareMathSymbol{\closeket}{\mathord}{symbols}{"69}
\DeclareMathOperator{\Rre}{Re}
\DeclareMathOperator{\Iim}{Im}
\newcommand{\ket}[1]{{| #1 \closeket}}
\newcommand{\bra}[1]{{\openbra #1 |}}
\newcommand{\aver}[1]{{\lang #1 \rang}}
\begin{document}
\begin{frontmatter}
\title{Quantum chaos and fractals with atoms in cavities}
\author{M. Uleysky, L. Kon'kov, S. Prants}
 \ead{prants@poi.dvo.ru}
\address{Laboratory of Nonlinear Dynamical Systems,
V.I.Il'ichev Pacific Oceanological Institute of the Russian Academy of Sciences,
690041 Vladivostok, Russia}

\begin{abstract}
We study the coupled translational, electronic, and field dynamics of the
combined system ``a two-level atom + a single-mode quantized field + a
standing-wave ideal cavity''. In the semiclassical approximation with a
point-like atom, interacting with the classical field, the dynamics is
described by the Heisenberg equations for the atomic and field expectation
values which are known to produce semiclassical chaos under appropriate
conditions. We derive Hamilton -- Schr\"odinger equations for probability
amplitudes and averaged position and momentum of a point-like atom interacting
with the quantized field in a standing-wave cavity. They constitute,
in general, an
infinite-dimensional set of equations with an infinite number of integrals of
motion which may be reduced to a dynamical system with four degrees of
freedom if the quantized field is supposed to be initially prepared in a Fock
state. This system is found to produce semiquantum chaos with positive values
of the maximal Lyapunov exponent. At exact resonance, the
semiquantum dynamics is regular. At large values of detuning $|\delta|\gg 1$,
the Rabi atomic oscillations are usually shallow, and the dynamics is
found to be almost regular. The Doppler -- Rabi resonance, deep Rabi
oscillations that may occur at any large value of $|\delta|$ to be equal to
$|\alpha p_0|$, is found numerically and described analytically (with $\alpha$
to be the normalized recoil frequency and $p_0$ the initial atomic
momentum). Two gedanken experiments are proposed to detect manifestations of
semiquantum chaos in real experiments. It is shown that in the chaotic regime
values of the population inversion $z_\text{out}$, measured with atoms after
transversing a cavity, are so sensitive to small changes in the initial
inversion $z_\text{in}$ that the probability of detecting any value of
$z_\text{out}$ in the admissible interval $[-1,\,1]$ becomes almost unity in a
short time. Chaotic wandering of a two-level atom in a quantized Fock field
is shown to be fractal. Fractal-like structures, typical with chaotic
scattering, are numerically found in the dependence of the time of exit of
atoms from the cavity on their initial momenta.
\end{abstract}
\end{frontmatter}

\section{Introduction}
The correspondence between quantum and classical worlds has been a subject of
much interest from the early days of quantum mechanics. The emergence of
classical dynamical chaos from quantum mechanics is one of the most debated
problem in this field \cite{Chi}. {\it Isolated and bounded\/} quantum
systems do not show sensitive dependence on initial conditions in the same
way as classical systems due to discreteness of quantum phase space and lacking of
notion of trajectories in quantum mechanics. The Schr\"odinger equation for
an isolated quantum system demonstrates only (quasi)periodic solutions even
if the classical counterpart of the quantum system under consideration would
be chaotic. However, it is valid only if  the quantum system is assumed to be
absolutely isolated from the surroundings.

In this paper, we study the temporal evolution of the strongly coupled
atom-field system consisting of a single two-level atom interacting with a
single mode of the quantized field in a standing-wave ideal cavity without
any leakage of photons. If the atom is assumed to be at rest the respective
Jaynes -- Cummings Hamiltonian is known to be integrable in the rotating-wave
approximation both under fully quantum and semiclassical descriptions
\cite{JC}. If the atom moves with a constant velocity (the Raman -- Nath
approximation) in the direction along which the coefficient of the atom-field
coupling is changed (say, periodically) the semiclassical evolution of the
expectation values of the atomic and field operators has been shown to be
chaotic \cite{PK97,PKK99,P02} with positive values of the maximal Lyapunov
exponent in  respective ranges of control parameters. The same
time-periodic Hamiltonian has been shown to generate quasiperiodic solutions
of the respective time-dependent Schr\"odinger equation for the probability
amplitudes \cite{KP02}. The Jaynes -- Cummings system possesses two degrees
of freedom, the electronic (internal) atomic one and the field one. In fact,
when emitting and absorbing photons, an atom not only changes its internal
state but its velocity is changed as well due to the photon recoil. It is a
pronouncing effect with cold atoms in a laser field (see for a review, for
example, \cite{R}). Taking into account the translational (external) atomic
degree of freedom, we get the autonomous Hamiltonian (\ref{1}) with three
degrees of freedom. The respective semiclassical equations of motion for the
expectation values of the atomic position and momentum operators, the atomic
population operator, and the combined atom-field operators have been shown to
be chaotic \cite{PK01,PS01} with positive values of the maximal Lyapunov
exponent. The semiclassical atom-field dynamics has been shown to demonstrate
many interesting features, including the interaction of nonlinear resonances \cite{P},
atomic fractals \cite{P02,PA}, L\'evy flights and anomalous atomic diffusion
\cite{PEZ}, and the Doppler -- Rabi resonance \cite{AP}.

In this paper, we go further in quantizing the atom-photon interaction
in a stan\-ding-wave cavity. The field and internal atomic degrees of freedoms
are treated as fully quantum ones obeying the Schr\"odinger equation. They are
coupled to the external atomic degree of freedom obeying the Hamilton
equations. Such a quantum-classical hybrid may be considered as a
reality-based model of interaction between a quantum system and the
surroundings. From the standpoint of dynamical system theory, the hybrid is
described by an infinite-dimensional set of nonlinear ODE's with an infinite
number of integrals of motion with clear physical meanings (see
Eqs.~(\ref{18}) -- (\ref{21})). Any state of the quantized field may be
represented as a superposition of a number of the so-called Fock
states $\ket{n}$, where $n$ is the number of photons in the respective state.
Any state of a two-level atom is a superposition of its ground and excited
states. The Hilbert space of the quantized atom-field subsystem (which should
be treated as a whole unity) is an infinite direct sum of two-dimensional
subspaces in each of which the so-called number of excitations should be
conserved in the process of evolution. The atom-field quantized
subsystem evolves in such a way that
transitions, belonging to the subspaces with different values of the number
of excitations, are forbidden. The main aim of the paper is to investigate
the effects of quantization on those properties of the system that produce
dynamical chaos.

This paper is organized as follows. In Sec.~\ref{two} we introduce the system
under consideration and the respective Hamiltonian. In Sec.~\ref{three} we
derive the Heisenberg equations for the expectation values of the atomic and
field operators and review briefly the properties of semiclassical chaos. Our
main results are given in Sec.~\ref{four} where we derive the Hamilton --
Schr\"odinger equations of motion for a two-level atom in a quantized field,
study the quantum Doppler -- Rabi resonance, the properties of semiquantum
chaos, and atomic fractals in the Fock quantized field.

\section{Two-level atom with recoil in a standing-wave cavity}\label{two}

We consider a single two-level atom with the frequency $\omega_a$ of an
electric dipole transition and mass $m_a$ moving in an ideal cavity which
sustains a single standing-wave mode along the axis $x$ with the frequency
$\omega_f$ and the  wave vector $k_f$. In the strong coupling limit, where the
coefficient of the coupling $\Omega_0$ is much greater than all the
relaxation rates, the atom-field dynamics may be treated as Hamiltonian with
the respective operator
\begin{equation}
\hat H=\frac{\hat P^2}{2m_a}+\frac{\hbar \omega_a}{2}\,\hat\sigma_z+
\hbar\omega_f\hat a^\dag\hat a-\hbar\Omega_0\left(\hat a^\dag\hat\sigma_-+
\hat a\hat\sigma_+\right)\,\cos{k_f\hat X},
\label{1}
\end{equation}
whose summands describe the kinetic and internal energies of the atom, the
field energy, and the energy of the atom-field interaction, respectively. The
momentum $\hat P$, position $\hat X$, atomic $\hat\sigma$, and field $\hat
a,\,\hat a^\dag$ operators satisfy the standard commutations relations:
\begin{equation}
[\hat X,\,\hat P]=i\hbar,\quad [\hat\sigma_\pm,\,\hat\sigma_z]=\mp
2\hat\sigma_\pm,\quad [\hat\sigma_+,\,\hat\sigma_-]=\hat\sigma_z,\quad
[\hat a,\,\hat a^\dag]=1.
\label{2}
\end{equation}
Operators belonging to different degrees of freedom commute with each other
at the same time moment.

In the process of emitting and absorbing photons an atom not only changes its
internal electronic state but its external translational state is changed as
well due to the photon recoil effect. An interplay between the electronic,
translational, and field degrees of freedom of the strongly coupled
atom-field system may be described as in the Heisenberg as in the
Schr\"odinger pictures.

\section{Heisenberg equations for the atomic and field expectation values and
semiclassical dynamics}\label{three}
It is convenient to write down the Heisenberg equations for the following
operators:
\begin{equation}
\hat x=k_f\hat X,\quad \hat p=\frac{\hat P}{\hbar k_f},\quad
\hat u=\frac{\hat a^\dag\hat\sigma_-+\hat a\hat\sigma_+}{\sqrt{\hat N}},\quad
\hat v=i\,\frac{\hat a^\dag\hat\sigma_--\hat a\hat\sigma_+}{\sqrt{\hat N}},
\quad \hat\sigma_z,
\label{3}
\end{equation}
where $\hat N=\hat a^\dag\hat a+(\hat\sigma_z+\hat I)/2$ is a constant
operator of the total number of excitations and $\hat I$ the identity
operator. The derivative of an arbitrary
operator $\hat A$ with respect to the normalized time $\tau=\Omega_0 t$
\begin{equation}
i\hbar\dot{\hat A}=[\hat A,\,\hat H]
\label{4}
\end{equation}
results in the following Heisenberg equations for the operators (\ref{3}):
\begin{equation}
\begin{aligned}
\dot{\hat x}\,&=\,\alpha\hat p,\vphantom{\sqrt{\hat N}}\\
\dot{\hat p}\,&=\,-\sqrt{\hat N}\,\hat u\sin\hat x,\\
\dot{\hat u}\,&=\,\delta\hat v,\vphantom{\sqrt{\hat N}}\\
\dot{\hat v}\,&=\,-\delta\hat u+2\sqrt{\hat N}\,\hat\sigma_z\cos\hat x,\\
\lefteqn{\dot{\hat \sigma}_{z}}\phantom{\dot{\hat \sigma}\,}
&=\,-2\sqrt{\hat N}\,\hat v\cos\hat x,
\end{aligned}
\label{5}
\end{equation}
where the control parameters
\begin{equation}
\alpha=\frac{\hbar k_f^2}{m_a\Omega_0},\qquad
\delta=\frac{\omega_f-\omega_a}{\Omega_0}
\label{6}
\end{equation}
are the normalized recoil frequency and the detuning between the field and
the atomic frequencies, respectively. The parameter
$\alpha=2\omega_R/\Omega_0$ characterizes the average change in translational
energy $\hbar\omega_R$ in the process of emission and absorption of a photon.
The set (\ref{5}) is not closed. In order to describe fully quantized
dynamics one should write down the equations of motion for all the products
of the operators (\ref{3}) and their functions which, in turn, would generate
another operator products and respective equations of motion and so on. This
process, in general, generates an infinite hierarchy of operator equations.
The simplest way to resolve this problem is to take quantum expectation
values over an initial quantum state and to factorize all the operator
products in Eqs.~(\ref{5}). In this way one gets the closed dynamical system
\begin{equation}
\begin{aligned}
\dot x&=\alpha p,\vphantom{\sqrt{N}}\\
\dot p&=-\sqrt{N}\,u\sin x,\\
\dot u&=\delta v,\vphantom{\sqrt{N}}\\
\dot v&=-\delta u+2\sqrt{N}\,z\cos x,\\
\dot z&=-2\sqrt{N}\,v\cos x,
\end{aligned}
\label{7}
\end{equation}
for the classical variables, namely, the atomic position $x=\aver{\hat x}$ and
momentum $p=\aver{\hat p}$, the atom-field variables $u=\aver{\hat u}$ and
$v=\aver{\hat v}$, and the atomic population inversion
$z=\aver{\hat\sigma_z}$. A conserved number of excitations in the system
\mbox{$N=\aver{\hat N}=n+(z+1)/2$} is the additional control parameter. The
set (\ref{7}) has two integrals of motion
\begin{equation}
W=\frac{\alpha}{2}\,p^2-u\sqrt{N}\cos x-\frac{\delta}{2}\,z,\qquad
R^2=u^2+v^2+z^2,
\label{8}
\end{equation}
where $W$ is the conserved total energy and $R^2$ reflects the conservation
of the length of the Bloch vector in the limit of the large number of
photons, $n=\aver{\hat a^\dag\hat a}\gg 1$. The set (\ref{7}) (with slightly
another normalization) has been derived in \cite{PK01,PS01}.

In deriving Eqs.~(\ref{7}) from an infinite hierarchy of Heisenberg operator
equations, we treat an atom as a point particle that may be justified if its
momentum is much greater then the photon momentum $\hbar k_f$, i. e. if
$|p|\gg 1$. Factorization of all the operator products in the respective
operator equations means that we do not take into account either quantum
nature of the field or quantum correlations between all the atomic and field
degrees of freedom. It is justified if $n\gg 1$. Moreover, the procedure of
reducing operator equations to semiclassical ones is not unique because the
form of the resulting semiclassical equations depends on the factorization
procedure. Semiclassical equations of motion, different from (\ref{7}), have
been derived in \cite{P02,PEZ} with the same Hamiltonian (\ref{1}). All of
them, of course, has the same form in the limit $n\gg 1$.

It has been found in a series of papers \cite{PK01,PS01,P02,P,PEZ} that
the semiclassical equations of motion of the strongly coupled atom-field
system in a standing wave cavity produce different types of motion. At exact
resonance, $\delta=0$, the motion is regular since the set (\ref{7})
gains an additional conserved quantity $u(\tau)=u(0)=u_0$ which reflects
the conservation of the atom-field interaction energy at $\delta=0$.
Depending on the values of the initial atomic momentum $p_0$, an atom either
oscillates periodically in a potential well of the standing wave or flies
over its potential hills. In resonance, the optical potential $U=-u_0\sqrt{N}
\cos x-\delta z/2$ coincides with the standing wave structure. The
oscillations of the internal atomic energy, the so-called Rabi oscillations
$z(\tau)$, and the oscillations of the atom-field variables, $u(\tau)$ and
$v(\tau)$, are regular as well. The center-of-mass motion of the atom does
not depend on the Rabi oscillations, but its frequency depends on the initial
interaction energy $u_0$ since it determines the depth of the optical
potential wells. In contrary, the Rabi oscillations depend on the
translational motion since the strength of the atom-field coupling depends on
the position of an atom in a cavity. The respective exact solutions of
Eqs.~(\ref{7}) with $\delta=0$ one can found in \cite{PS01}.

Out of resonance, $\delta\ne 0$, the set (\ref{7}) with two integrals
(\ref{8})  is an autonomous Hamiltonian system whose motion takes place on a
three-dimensional hypersurface. It has, generally speaking, a positive
Lyapunov exponent $\lambda$ which has been computed in the paper \cite{PK01}
as a function of the control parameters $\alpha$ and $\delta$ and of the
initial atomic momentum $p_0$. It follows from the first two equations in
(\ref{7}) that the atomic center-of-mass motion is described by the equation
for a physical pendulum with a frequency modulation
\begin{equation}
\ddot x+\alpha\sqrt{N}\,u(\tau)\sin x=0,
\label{9}
\end{equation}
where $u$ is a function of time and all the other variables. Following to
\cite{Z,P} one can show that Eq.~(\ref{9}) may produce a stochastic layer
in a neighbourhood of the unperturped separatrix. Analogously to what has
been done in \cite{P}, the normalized (to a separatrix value) width of the
stochastic layer may be estimated as follows
\begin{equation}
\Delta\simeq 8\pi\left(\Omega/\omega\right)^3
\exp\left(-\frac{\pi\Omega}{2\omega}\right),
\label{10}
\end{equation}
where $\Omega=\sqrt{\delta^2+4N}$ is the normalized Rabi frequency and
the frequency $\omega=\sqrt{2\alpha N^{3/2}|\delta|}/\Omega$ characterizes
small-amplitude oscillations. It should be noted that $\Delta$ gives the
lower bound for the layer width (see \cite{P}).

Center-of-mass motion of a two-level atom in an ideal standing-wave cavity
has been found in \cite{PEZ} to be anomalous. A typical chaotic atomic
trajectory consists of intervals of regular motion with an almost constant
velocity in each interval (L\'evy flights) interrupted by erratic walks. Such
an intermittency is typical of Hamiltonian systems with nonhomogeneous phase
space with a fractal-like structure consisting of KAM tori, cantori, chains of
islands, stochastic sea, etc. \cite{Za}. From this point of view, the L\'evy
flights may be understood as those trajectories that ``stick'' to islands
boundaries for a long time. The representative point on a typical chaotic
trajectory sooner or later approaches, as closely as desirable, an island
boundary that separates regular and chaotic motions. Nearby such a boundary,
the maximal Lyapunov exponent $\lambda$ goes to zero, and cantories block the
trajectory escape to the stochastic sea. As a result, the atomic motion is
almost regular for a time that may be very long. From the physical point of
view, the intermittency of the center-of-mass motion is due to the
intermittent oscillations of the effective optical potential
$U_{\rm eff}=-\sqrt{N}\,u\cos x-\delta z/2$ which governs the translational motion of
the atom \cite{AP}.

The L\'evy flights impact the statistical properties of the atomic motion
resulting in the anomalous diffusion. It was numerically found in
\cite{PEZ} that the second moment of the position of atom in a cavity
evolves in time as $\bar x^2\sim\tau^\mu$, where the transport exponent $\mu$
may vary from the value $\mu\simeq 1$, corresponding to the normal diffusion,
to the value $\mu\simeq 2$ corresponding to the superdiffusion. The
Poincar\'e  theorem states that every trajectory of a closed conservative
dynamical system, except for trajectories of the set of zero measure, returns
arbitrarily close to its origin infinitely many times. The recurrence time
distribution in the system with perfect mixing is known to be Poissonian,
$P(\tau)=h^{-1}e^{-h\tau}$, where $h$ is the Kolmogorov-Sinai entropy. The
motion with intermittency and L\'evy flights leads to the power law,
$P(\tau)\sim \tau^{-\gamma}$ at $\tau\to\infty$. The exponents $\mu$ and
$\gamma$ are related to each other, and their values depends on the values of
the control parameters $\alpha$, $\delta$, and $N$ because by changing them
one changes the topology of the phase space.

The atom-photon interaction in a cavity may be considered as the chaotic
scattering problem \cite{P02,PA} where a two-level atom is scattered by
the standing-wave light or by an optical potential. In difference from the
usual scattering of atoms by light \cite{R}, the effective optical potential
$U_{\rm eff}$ of the strongly coupled atom-field system depends not only on the field
but on the atomic variables as well. Let two atomic detectors to be placed at
the cavity mirrors and let them detect the time $T$ of atomic exit from the
cavity. Let the identically prepared atoms with given initial momenta $p_0$
be placed one by one in the middle of the cavity. The dependence $T(p_0)$ has
been found in \cite{P02,PA} to have a beautiful selfsimilar structure with
the Hausdorff dimension to be equal to $d\simeq 1.84$. Tiny interplay between
all the degrees of freedom is responsible for trapping atoms with
$T\to\infty$ even in a very short microcavity. Simulation in the cited papers
has been performed with a cavity whose length is equal to two standing-wave
lengths. Two kinds of atomic fractals have been found in \cite{PA}, a
countable fractal (a set of $p_0$ generating separatrix-like atomic
trajectories) and a seemingly uncountable fractal with a set of $p_0$
generating infinite walkings of atoms inside the cavity.

\section{Hamilton-Schr\"odinger equations and semiquantum
dynamics}\label{four}
\subsection{Derivation of equations of motion}
In this section we again consider a two-level atom as a point particle but
electronic and field degrees of freedom are treated as fully quantum ones in
the Schr\"odinger picture. It enables us to study the role that field quantum
statistic and atomic superposition play in the full atom-field dynamics
including chaos. We start with the Hamiltonian $\hat H$ (\ref{1}).

The Hamilton equations for the classical translational degree of freedom is
easily found
\begin{equation}
\frac{d\aver{\hat X}}{dt}=
\frac{\partial\aver{\hat H}}{\partial\aver{\hat P}},\qquad
\frac{d\aver{\hat P}}{dt}=
-\frac{\partial\aver{\hat H}}{\partial\aver{\hat X}},
\label{11}
\end{equation}
where $\aver{\dots}$ denotes an expectation value of the corresponding
operator over a quantum state $\ket{\Psi}$ of the electronic-field
Hamiltonian. Using the same normalizations and notations as in the preceding
section, we get
\begin{equation}
\dot x=\alpha p,\qquad\dot p=-\aver{\hat u}\sin x,
\label{12}
\end{equation}
where $\aver{\hat u}=\bra{\Psi(\tau)}\hat u_0\ket{\Psi(\tau)}$.

Let us expand a state vector of the electronic-field subsystem over the basic
energetic atomic states $\ket{2}$ and $\ket{1}$ and the Fock field states
$\ket{n}$
\begin{equation}
\ket{\Psi(\tau)}=\sum_{n=0}^\infty\Bigl(
a_n(\tau)\ket{2,\,n}+b_n(\tau)\ket{1,\,n}\Bigr),
\label{13}
\end{equation}
where $a_n$ and $b_n$ are the probability amplitudes to find the atom in its
excited or ground state with $n$ photons in the mode, respectively.
Substitution of the vector (\ref{13}) in the time-dependent Schr\"odinger
equation
\begin{equation}
i\hbar\ket{\dot\Psi}=\hat H\ket{\Psi}
\label{14}
\end{equation}
gives the infinite-dimensional set of the coupled ODE's
\begin{equation}
\begin{aligned}
\dot a_n&=-i\left(\Delta_a a_n-\sqrt{n+1}\, b_{n+1} \cos x\right),&\\
\dot b_{n+1}&=
\phantom{-}
i\left(\Delta_b b_{n+1}^* -\sqrt{n+1}\, a_n^* \cos x\right),&
\quad n=0,\,1,\,2,\,\dots,
\end{aligned}
\label{15}
\end{equation}
where $\Delta_a=n\omega_f+\omega_a/2$ and
$\Delta_b=(n+1)\omega_f-\omega_a/2$. Introducing the following combinations
of the probability amplitudes:
\begin{equation}
u_n=2\Rre{\left(a_nb_{n+1}^* \right)},\quad
v_n=-2\Iim{\left(a_nb_{n+1}^*\right)},\quad
z_n=|a_n|^2-|b_{n+1}|^2,
\label{16}
\end{equation}
we get the quantum Bloch-like equations
\begin{equation}
\begin{aligned}
\dot u_n&=\delta v_n,\\
\dot v_n&=-\delta u_n+2\sqrt{n+1}\,z_n\cos x,\\
\dot z_n&=-2\sqrt{n+1}\,v_n\cos x,\qquad\qquad\quad n=0,\,1,\,2,\,\dots,
\end{aligned}
\label{17}
\end{equation}
where the detuning $\delta$ is the same as in (\ref{6}). After computing
$\aver{\hat u}$,
we obtain our basic Hamilton-Schr\"odinger equations
\begin{equation}
\begin{aligned}
\dot x_{\phantom n}&=\alpha p,\\
\dot p_{\phantom n}&=
-{\textstyle\sum\limits_{n=0}^\infty}
\sqrt{n+1}\, u_n\sin x,\\
\dot u_n&=\delta v_n,\\
\dot v_n&=-\delta u_n+2\sqrt{n+1}\, z_n\cos x,\\
\dot z_n&=-2 \sqrt{n+1}\, v_n\cos x, \qquad\qquad\quad n=0,\,1,\,2,\,\dots.
\end{aligned}
\label{18}
\end{equation}
This infinite set of {\it nonlinear} ODE's possesses an infinite number of
the integrals of motion, the total energy integral
\begin{equation}
W=\frac{\alpha}{2}\,p^2-\sum_{n=0}^\infty\sqrt{n+1}\,u_n\cos x -
\frac{\delta}{2}\sum_{n=0}^\infty z_n,
\label{19}
\end{equation}
the Bloch-like integrals for each $n$
\begin{equation}
R_n^2=u_n^2+v_n^2+z_n^2,
\label{20}
\end{equation}
and the global integral reflecting conservation of the total probability
\begin{equation}
\sum_{n=0}^\infty R_n=1.
\label{21}
\end{equation}
The quantities that can be measured in real experiments are the atomic
position $x$, momentum $p$ and the atomic population inversion
\begin{equation}
z(\tau)=\sum_{n=0}^\infty z_n(\tau).
\label{22}
\end{equation}

The semiquantum equations (\ref{18}) --- (\ref{21}) should be compared with
the semiclassical equations (\ref{7}) and (\ref{8}) because they describe the
same physical situation but on the different ground. They are exactly
identical in the case of the initial Fock state of the cavity field with
$\bar n$ quanta and the atom to be prepared initially in one of its energetic
states. It is easy to show that an infinite number of equations (\ref{18})
reduces in this case to five equations (\ref{7}) with
$N=\bar n+(z+1)/2=\bar n+1$ if $\ket{\Psi(0)}=\ket{2,\,n}$ and $N=\bar n$ if
$\ket{\Psi(0)}=\ket{1,\,n}$. If the atom is initially prepared in a general
superposition state and the field is in the Fock state $\ket{n}$
\begin{equation}
\ket{\Psi(0)}=a_n(0)\ket{2,\,n}+b_n(0)\ket{1,\,n},\qquad
\bigl|a_n(0)\bigr|^2+\bigl|b_n(0)\bigr|^2=1,
\label{23}
\end{equation}
we get from Eqs.~(\ref{18})
\begin{equation}
\begin{gathered}
\begin{aligned}
\dot x&=\alpha p,\\
\dot p&=-\left(\sqrt{\bar n}\,u_{n-1}+\sqrt{\bar n+1}\,u_n\right)\sin x,
\end{aligned}\\
\begin{aligned}
\dot u_{n-1}&=\delta v_{n-1},&
\dot u_n&=\delta v_n,\\
\dot v_{n-1}&=-\delta u_{n-1}+2\sqrt{\bar n}\,z_{n-1}\cos x,\quad&
\dot v_n&=-\delta u_n+2\sqrt{\bar n+1}\,z_n\cos x,\\
\dot z_{n-1}&=-2\sqrt{\bar n}\,v_{n-1}\cos x,&
\dot z_n&=-2\sqrt{\bar n+1}\,v_n\cos x
\end{aligned}
\end{gathered}
\label{24}
\end{equation}
with the respective integrals
\begin{equation}
\begin{aligned}
&W=\frac{\alpha}{2}\,p^2-\left(\sqrt{\bar n}\,u_{n-1}+
\sqrt{\bar n+1}\,u_n\right)\cos x-\frac{\delta}{2}(z_{n-1}+z_n),\\
&R_{n-1}^2=u_{n-1}^2+v_{n-1}^2+z_{n-1}^2=\bigl|b_n(0)\bigr|^4,\\
&R_n^2=u_n^2+v_n^2+z_n^2=\bigl|a_n(0)\bigr|^4,\\
&R_{n-1}+R_n=1,
\end{aligned}
\label{25}
\end{equation}
and initial conditions
\begin{multline}
x(0)=x_0,\quad p(0)=p_0,\quad
z_{n-1}(0)=-\bigl|b_n(0)\bigr|^2,\quad z_n(0)=\bigl|a_n(0)\bigr|^2,\\
u_{n-1}(0)=u_n(0)=v_{n-1}(0)=v_n(0)=0.
\label{26}
\end{multline}

\subsection{Doppler -- Rabi resonance}

If the field frequency $\omega_f$ is far detuned from the frequency of the
atomic working transition $\omega_a$, i.~e.\ if $|\delta|\gg 1$, then the
Rabi oscillations $z(\tau)$ should be very shallow as oscillations of an
oscillator under influence of a far-detuned time-dependent force. It is
correct if the atom does not move. The Doppler effect with a moving atom
causes an interesting effect of the Doppler -- Rabi resonance if the
condition $|\alpha p|\simeq|\delta|$ is fulfilled. A standing wave is a sum
of two running waves moving in the opposite directions. In the reference
frame of the moving atom, their frequencies are different due to the Doppler
effect
\begin{equation}
\omega_1=\omega_f-\frac{v_a}{c}\,\omega_f,\qquad
\omega_2=\omega_f+\frac{v_a}{c}\,\omega_f,
\label{27}
\end{equation}
where $v_a$ and $c$ are the atomic and light velocities, respectively, and
$v_a\ll c$. If the initial atomic momentum $p_0$ is sufficiently large, the
Raman -- Nath condition, $p\simeq p_0$ is valid. Let us define dimensionless
detunings between the frequencies of the atomic transition and the running
waves as follows:
\begin{equation}
\delta_1=\frac{\omega_1-\omega_a}{\Omega_0}=\delta-\alpha p_0,\qquad
\delta_2=\frac{\omega_2-\omega_a}{\Omega_0}=\delta+\alpha p_0.
\label{28}
\end{equation}
It is  evident that the atom comes in resonance with one of the running waves
if $|\alpha p_0|\simeq|\delta|$. If $|\delta|\gg 1$, the interaction of the
atom with the other running wave is negligibly small. Suppose that the atom
is initially prepared in the ground state, i.~e.\ $z_{n-1}(0)=-1$,
$z_n(0)=0$, and $z(0)=-1$, than the equations of motion (\ref{24}) for the
atom with a constant speed and with the other initial conditions written in
(\ref{26}) are reduced to the following simple set:
\begin{equation}
\begin{aligned}
\dot u_n&=\delta v_n,\\
\dot v_n&=-\delta u_n+\sqrt{\bar n+1}\,z_n,\\
\dot z_n&=-\sqrt{\bar n+1}\,v_n.
\end{aligned}
\label{29}
\end{equation}
It should be noted that the amplitude of the running wave is half of the
standing-wave amplitude, and the atom-field interaction does not depend on
atomic position. The solution for the atomic population inversion with given
initial conditions is easy to find
\begin{equation}
z(\tau)=z_n(\tau)=-\left(\frac{\delta-\alpha p_0}{\Omega_n}\right)^2-
\frac{\sqrt{\bar n+1}}{\Omega_n^2}\cos \Omega_n\tau,
\label{30}
\end{equation}
where $\Omega_n=\sqrt{(\delta-\alpha p_0)^2+\sqrt{\bar n+1}}$ is
the Rabi frequency. In particular, at exact Doppler -- Rabi resonance,
$|\alpha p_0|=|\delta|$, the inversion oscillates at the frequency $(\bar
n+1)^{1/4}$ and its amplitude is maximal.

\begin{figure}[h!t]
\begin{center}
\includegraphics[width=0.8\textwidth]{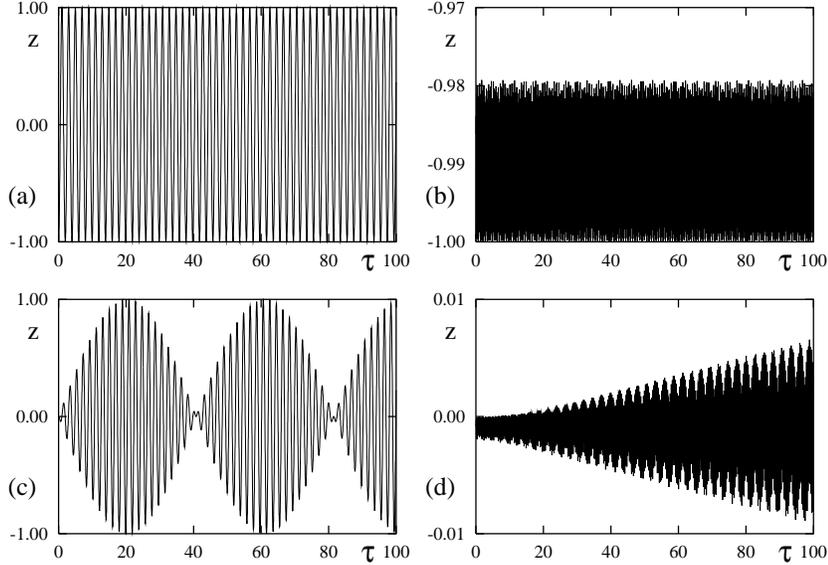}
\caption{Doppler -- Rabi resonance at the condition $\alpha p_0=\delta$ with $\delta=32$,
$p_0=32000$, $\alpha=0.001$, (a) $z(0)=-1$ and (c) $z(0)=0$. Shallow Rabi oscillations
with the same value of $p_0$ but out off the resonance are shown for comparison with
(b) $z(0)=-1$, $\delta=1$ and (d) $z(0)=0$, $\delta=10$.}\label{fig1}
\end{center}
\end{figure}
The Doppler -- Rabi resonance is illustrated in Fig.~\ref{fig1} where we plot
the Rabi oscillation signals $z(\tau)$ computed with the full set (\ref{24})
and two initial states of the atom when it is prepared in the ground level
with $z(0)=-1$ and in the superposition state with $z(0)=0$
($z_{n-1}(0)=-1/2$, and $z_n(0)=1/2$). In our simulation $\alpha=10^{-3}$,
$\bar n=10$ and $x_0=0$. When $\delta=32$ and $p_0=32000$ and the resonance
condition is fulfilled we really see the maximal Rabi oscillations with both
the initial conditions. The signal in Fig.~\ref{fig1}a with $z(0)=-1$ is
described by the analytical solution (\ref{30}). When $z(0)=0$ we have two
oscillators with different frequencies (see (\ref{24})) and the resulting
signal $z(\tau)=z_{n-1}(\tau)+z_n(\tau)$ in Fig.~\ref{fig1}c demonstrates a
beating with two harmonic components of the type of (\ref{30}) with
$\Omega_n=(\bar n+1)^{1/4}$ and $\Omega_{n-1}=\bar n^{1/4}$. For comparison,
we plot in Fig.~\ref{fig1}b and d very shallow Rabi oscillations with the
same value of the atomic momentum $p_0=32000$ but out off the resonance,
$|\alpha p_0|\ne|\delta|$. The amplitude of the Rabi oscillations in
Fig.~\ref{fig1}b with $z(0)=-1$ and $\delta=1$ does not exceed $2\%$ of the
amplitude of the resonant oscillations in Fig.~\ref{fig1}a. The same is valid
with another initial value of the atomic inversion population, $z(0)=0$
(compare, please, Figs.~\ref{fig1}c and d).

In conclusion, deep Rabi oscillations are possible at as large values of the
detuning $|\delta|$ as desirable if a two-level atom moves with the
corresponding velocity.

\subsection{Chaos in a quantized Fock field}

Lyapunov exponents are known to be quantitative indicators of chaos in
dynamical systems. They characterize the behavior of close trajectories in
phase space. If the quantized field is initially prepared in a Fock state
$\ket{n}$ with $n$ quanta in the mode, the Hamilton -- Schr\"odinger
equations of motion have the form (\ref{24}) for an arbitrary internal atomic
state. The 8-dimensional dynamical system (\ref{24}) with 4 integrals of
motion (\ref{25}) has, as maximum, four nonzero Lyapunov exponents
$\lambda_i$ ($i=1,\,2,\,3,\,4$)
\begin{equation}
\lambda_i=\lim_{\tau\to\infty}\lambda_i(\tau),\qquad
\lambda_i(\tau)=\lim_{\Delta_i(0)\to 0}
\frac{1}{\tau}\ln\frac{\Delta_i(\tau)}{\Delta_i(0)},
\label{31}
\end{equation}
where $\Delta_i(\tau)$ is the distance (in the Euclidean sense) in the $i$-th
direction at the moment $\tau$ between two trajectories that were close to
each other at $\tau=0$. In Hamiltonian systems, due to the phase space volume
conservation, $\lambda_1+\lambda_2+\lambda_3+\lambda_4=0$, and
$\lambda_1=-\lambda_2$, $\lambda_3=-\lambda_4$. Computing the maximal
Lyapunov exponent $\lambda$, one measures an averaged rate of separation of
initially close trajectories.

\begin{figure}[h!t]
\begin{center}
\includegraphics[width=0.8\textwidth]{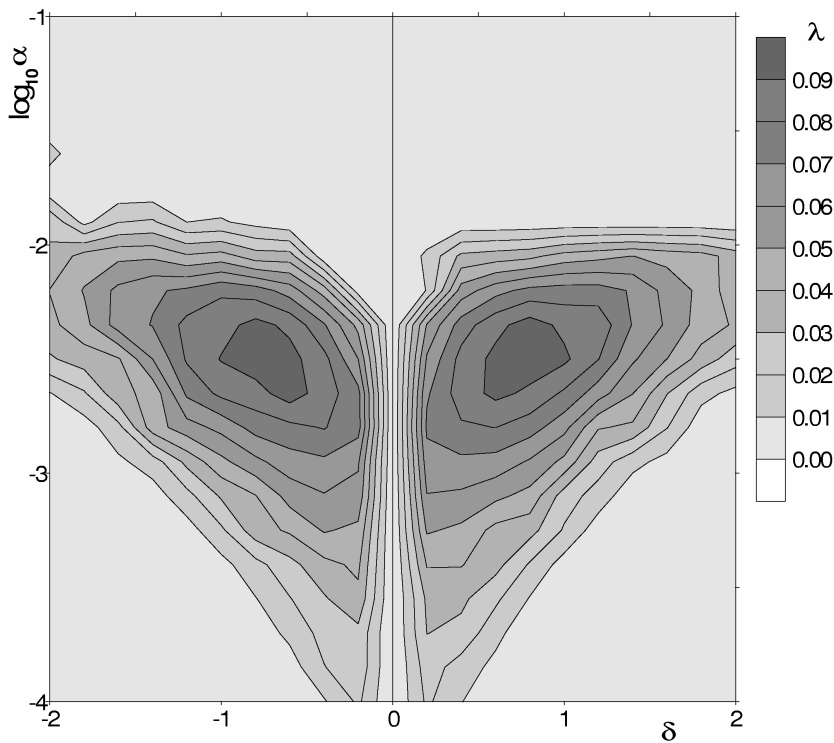}
\caption{The maximal Lyapunov exponent $\lambda$ with zero initial atomic population
inversion $z(0)=0$
and $\bar n=10$ photons in the Fock quantized field versus the atom-field detuning
$\delta$ and the logarithm of the dimensionless recoil frequency
$\alpha$.}\label{fig2}
\end{center}
\end{figure}
\begin{figure}[h!t]
\begin{center}
\includegraphics[width=0.8\textwidth]{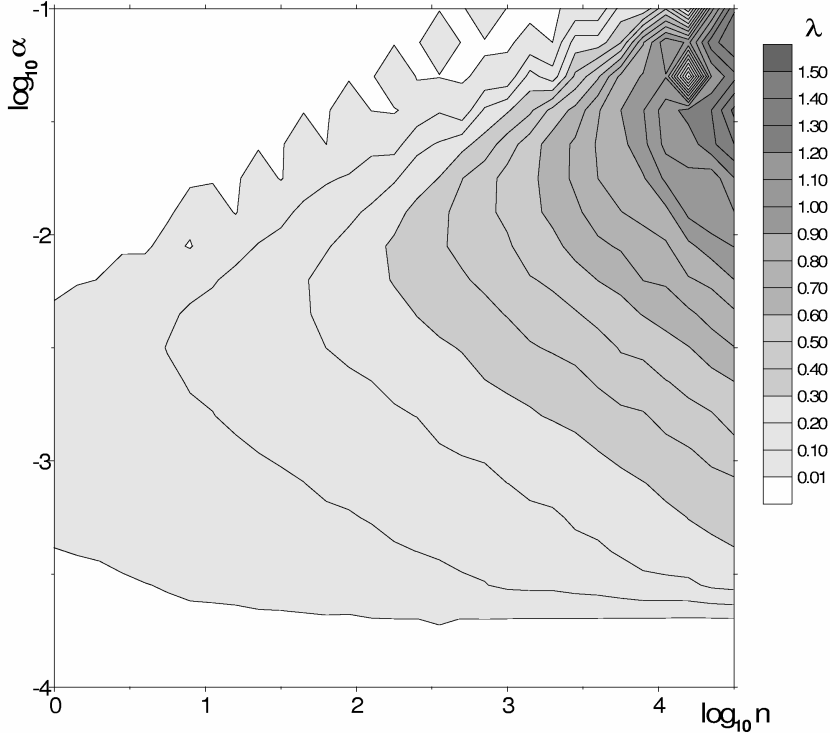}
\caption{The double logarithmic plot of $\lambda$ with an initially excited atom
$z(0)=1$ and $\bar n=10$ photons in the Fock quantized field versus $\alpha$ and $\bar n$
with $\delta=0.5$.}\label{fig3}
\end{center}
\end{figure}
The system has three control parameters, the initial number of photons in the
mode $\bar n$, the normalized recoil frequency $\alpha$ and the detuning
$\delta$, each of which may vary in a wide range of values. To diagnose chaos
it is instructive to compute the so-called topographic $\lambda$-maps
\cite{PK97,PKK99,IKP} which show by color modulation values of $\lambda$ in
the ranges of values of two control parameters with the other to be fixed.
When computing the $\lambda$-maps, we choose the following initial conditions: $x(0)=0$,
$p(0)=50$, $u_{n-1}(0)=u_n(0)=v_{n-1}(0)=v_n(0)=0$, $z_{n-1}(0)=-1/2$,
$z_n(0)=1/2$ and $z_{n-1}(0)=0$, $z_n(0)=1$. It means that the atom is
prepared initially in the state with zero population inversion $z(0)=0$ or
in the excited state, $z(0)=1$.  In Fig.~\ref{fig2} we show the $\lambda$-map
with $z(0)=0$ in the ranges of the detuning $|\delta|\leqslant 2$ and the
recoil frequency $-4\leqslant \log_{10}\alpha\leqslant -1$ with the fixed
value of the initial number of photons $\bar n=10$. The set (\ref{24}) is
integrable at exact atom-field resonance ($\delta=0$) with $\lambda=0$.  The
$\alpha$--$\delta$ map confirms this conjecture. The values of $\alpha\sim
10^{-4}\div 10^{-2}$, that correspond to positive values of $\lambda$, are
reasonable with real atoms \cite{PS01}, and one may expect chaotic atomic
motion in the respective ranges of $\alpha$ and $\delta$. Another
$\lambda$-map with $z(0)=1$ shows in Fig.~\ref{fig3} the value of $\lambda$
in dependence on $\alpha$ and $\bar n$ at $\delta=0.5$ in double logarithmic
scale. The magnitude of $\lambda$ grows, in average, with increasing the
initial number of photons in the cavity mode.

\begin{figure}[h!t]
\begin{center}
\parbox{0.49\textwidth}{\centerline{\large\bf (a)}\vspace{1mm}
\includegraphics[width=0.49\textwidth]{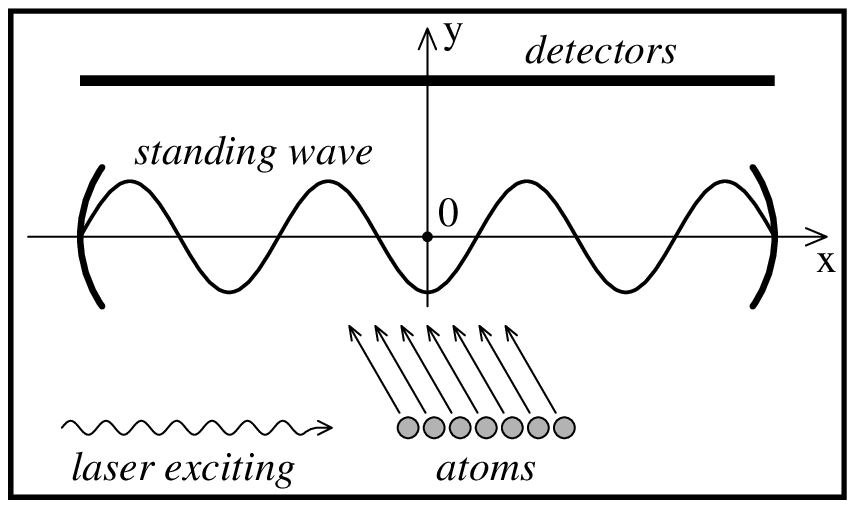}}
\hfil
\parbox{0.49\textwidth}{\centerline{\large\bf (b)}\vspace{1mm}
\includegraphics[width=0.49\textwidth]{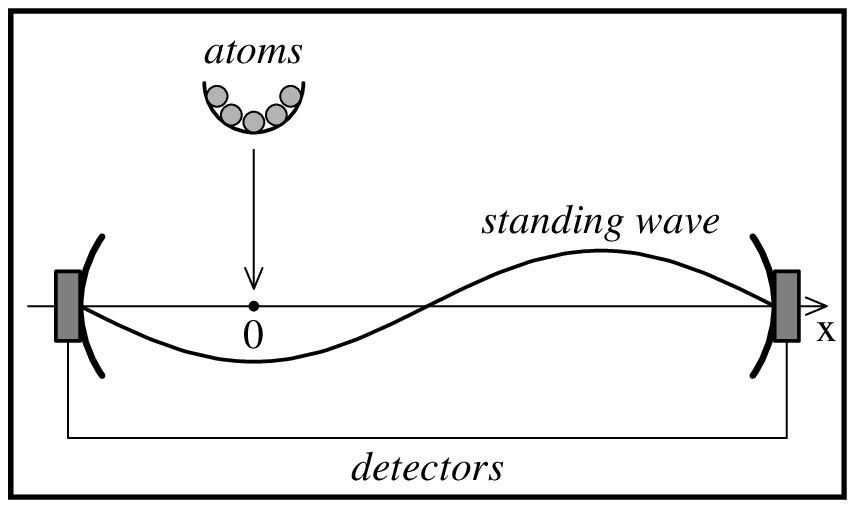}}
\caption{Schematic diagrams showing scattering of atoms at the standing
wave. (a) A monokinetic beam of atoms propagates transversally to the cavity
axis $x$, and the atoms are detected outside the cavity. (b) Atoms are placed
one by one inside the cavity with almost zero transversal velocity and are
detected at the cavity mirrors.}\label{fig4}
\end{center}
\end{figure}
A feasible scheme for detecting manifestation of chaos with hot two-level
Rydberg atoms moving in a high-Q microwave cavity has been proposed in
\cite{P02}. The same idea could be realized with cold usual atoms in a high-Q
microcavity. Consider a 2D-geometry of a gedanken experiment shown in
Fig.~\ref{fig4}a, where a monokinetic atomic beam propagates almost
perpendicularly to the cavity axis $x$. In a reference frame moving with a
constant velocity in the $y$-direction, there remains only the transverse
atomic motion along the axis $x$. One measures atomic population inversion
after passing the interaction zone. Before injecting atoms in the cavity, it
is necessary to prepare all the atoms in the same electronic state, say, in
the excited state, with the help of a $\pi$-pulse of the laser radiation. It
may be done with {\it only a finite accuracy}, say, equal to $\Delta
z_\text{in}$ for the initial population inversion $z_\text{in}$. The values
of the population inversion $z_\text{out}$ are measured with detectors at a
fixed time moment. If we would work with the values of the control parameters
corresponding to the regular atom-field dynamics, we would expect to have a
regular curve $z_\text{out}$--$z_\text{in}$. In the chaotic regime,  the
atomic inversion at the output can be predicted (within a certain confidence
interval $\Delta z$)  for a time not exceeding the so-called predictability
horizon
\begin{equation}
\tau_p\simeq\frac{1}{\lambda}\ln\frac{\Delta z}{\Delta z_\text{in}},
\label{predtime}
\end{equation}
which depends weakly on $\Delta z_\text{in}$ and $\Delta z$. Since the
maximal confidence interval lies in the range $|\Delta z|\leqslant 1$ and
$\lambda$ may reach the values of the order of $1.5$ (see $\lambda$-maps),
the predictability horizon in accordance with the formula (\ref{predtime})
can be very short: with $\lambda$ =0.5 $\tau_p$ may be of the order of 10 in units of the
reciprocal of the vacuum Rabi frequency $\Omega_0$ that correspond to
$t_p\simeq 10^{-7}$ s with the realistic value of $\Omega_0\simeq 10^8$
rad$\cdot$ s${}^{-1}$ \cite{HL}.

\begin{figure}[h!t]
\begin{center}
\includegraphics[width=0.8\textwidth]{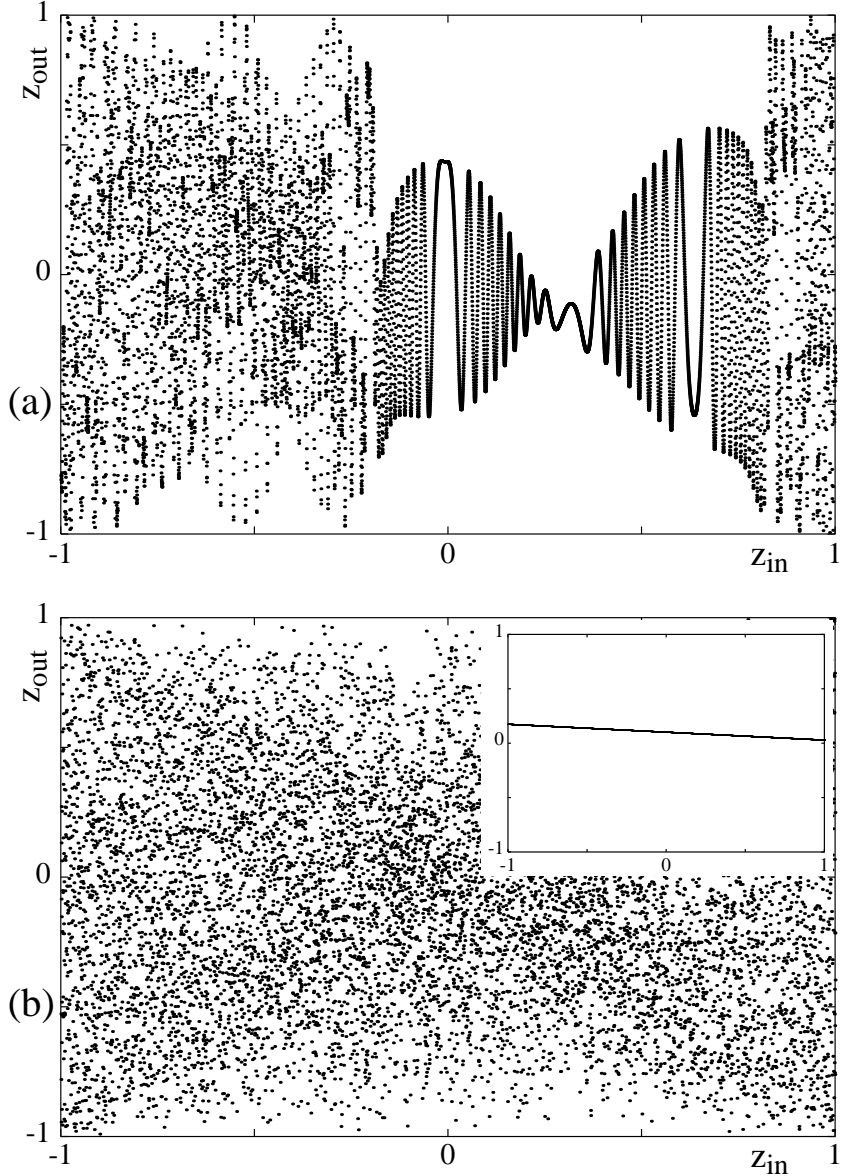}
\caption{Dependence of the output values of the atomic population inversion
$z_\text{out}$ on its initial values $z_\text{in}$ with $\delta=0.4$ (a)
at $\tau=100$ and (b) at $\tau=200$ with the inset showing this dependence
at the exact atom-field resonance, $\delta=0$.}\label{fig5}
\end{center}
\end{figure}
In the regular regime, the inevitable errors in preparing
$\Delta z_\text{in}$ produce the output errors $\Delta z_\text{out}$ of the
same order. In the chaotic regime, the initial uncertainty increases
exponentially resulting in a complete uncertainty of the detected population
inversion in a reasonable time. It is demonstrated in Fig.~\ref{fig5}, where
we plot the dependence of the values of $z(\tau)=z_\text{out}$ at $\tau=100$
(Fig.~\ref{fig5}a) and $\tau=200$ (Fig.~\ref{fig5}b) on the values of
$z(0)=z_\text{in}$ in the chaotic regime with $\delta=0.4$ and
$\lambda\simeq0.05$. It is evident from Fig.~\ref{fig5}a, that at the
detection time moment $\tau=100$ an initial error $\Delta
z_\text{in}=10^{-4}$ leads to a complete uncertainty $\Delta
z_\text{out}\simeq 2$ in rather large vicinities of the initial inversion
$z_\text{in}\simeq\pm 1$, whereas the dependence
$z_\text{out}$--$z_\text{in}$ is a regular one in the vicinity of
$z_\text{in}\simeq 0$. With increasing the detection time (see
Fig.~\ref{fig5}b at $\tau=200$), the probability of detecting any value of
$z_\text{out}$ in the interval $[-1,\,1]$ is almost unity in the whole
range of $z_\text{in}$. The formula (\ref{predtime}) gives the value of the
predictability horizon $\tau_p\simeq 200$ with $\lambda\simeq 0.05$. To feel
the difference, it is desirable to carry out a control experiment at the
exact resonance ($\delta=0$) when the atomic motion is fully regular with any
initial values. The dependence $z_\text{out}$--$z_\text{in}$ with $\delta=0$
is demonstrated in the inset in Fig.~\ref{fig5}b with all the other control
parameters and initial values being the same.

\subsection{Atomic fractals in a quantized field}

In this section, we treat the atom-photon interaction in a high-Q cavity as a
chaotic scattering problem \cite{Ott,BL}. Atoms from outside are injected
into a 1D-cavity, interact for a while with the cavity field and then they are
removed from the cavity by hook or by crook. Following to \cite{P02,PA}, let
us consider the scheme of scattering of atoms by the standing wave shown in
Fig.~\ref{fig4}b. Atoms, one by one, are placed at the point $x=0$ with
different initial values of the momentum $p_0$ along the cavity axis. For
simplicity, we suppose that they have no momentum in the other directions
(1D-geometry). We compute the time the atoms need to reach one of the
detectors placed at the cavity mirrors. The dependence of this exit time $T$
on the initial atomic momentum $p_0$ is studied under the other initial
conditions and parameters being the same.
To avoid complications that are not essential to the
main theme of this section, we consider the cavity with only two standing-wave
lengths. Before injecting into a cavity, atoms are suppose to be prepared in
the superposition state with $u_{n-1}(0)= u_{n}(0)=v_{n-1}(0)=v_{n}(0)=0$,
$z_{n-1}(0)=-1/2, z_{n}(0)=1/2$, i.~e.\ in the state with zero population
inversion $z(0)=z_{n-1}(0)+z_{n}(0)=0$.

\begin{figure}[h!t]
\begin{center}
{\bf (a)} \parbox[t]{0.6\textwidth}{\rule{0pt}{0pt}\\
\includegraphics[width=0.6\textwidth]{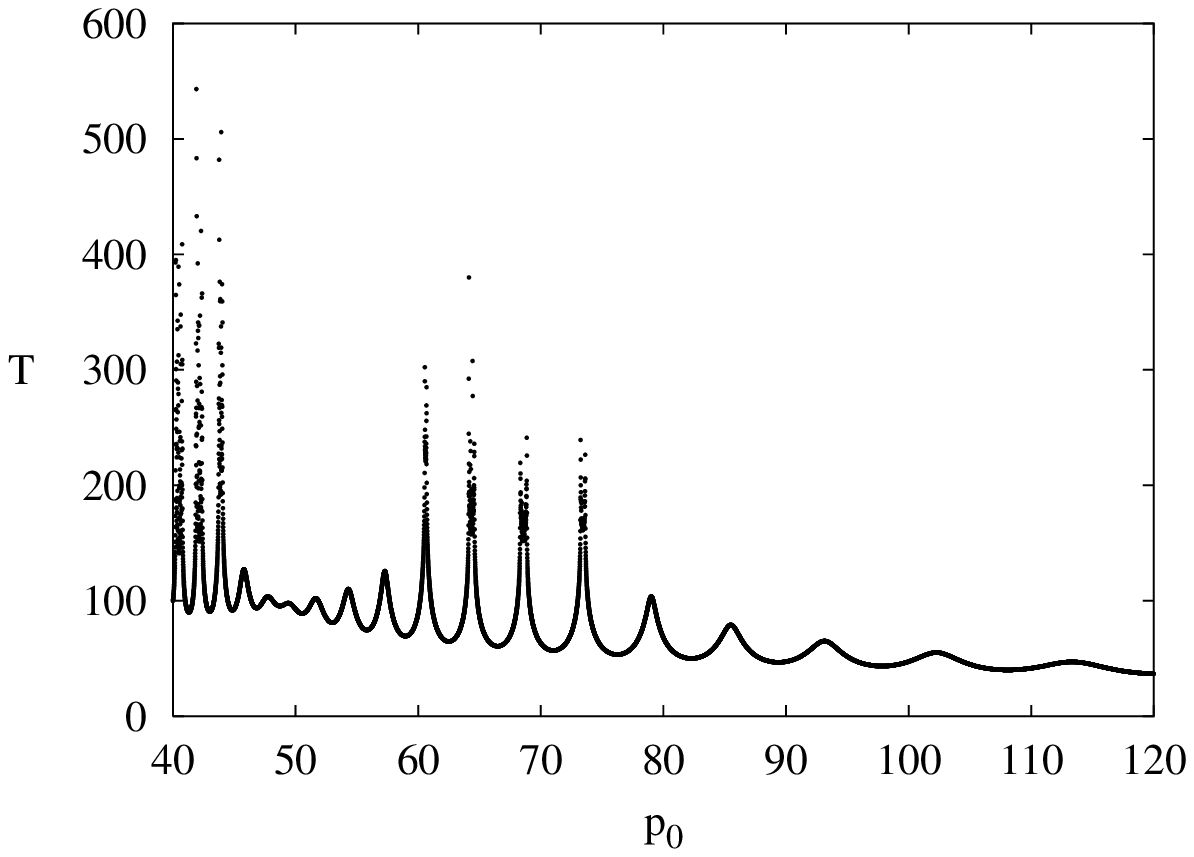}}\\
{\bf (b)} \parbox[t]{0.6\textwidth}{\rule{0pt}{0pt}\\
\includegraphics[width=0.6\textwidth]{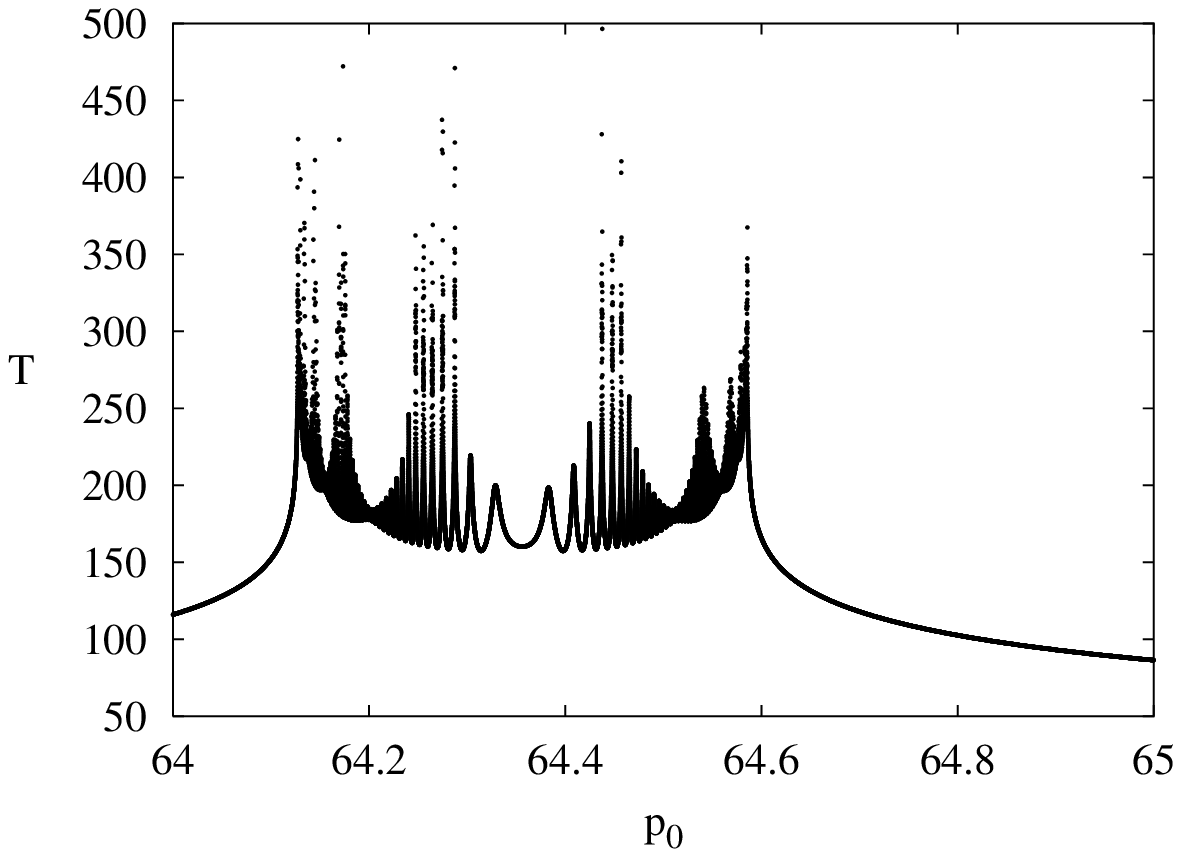}}\\
{\bf (c)} \parbox[t]{0.6\textwidth}{\rule{0pt}{0pt}\\
\includegraphics[width=0.6\textwidth]{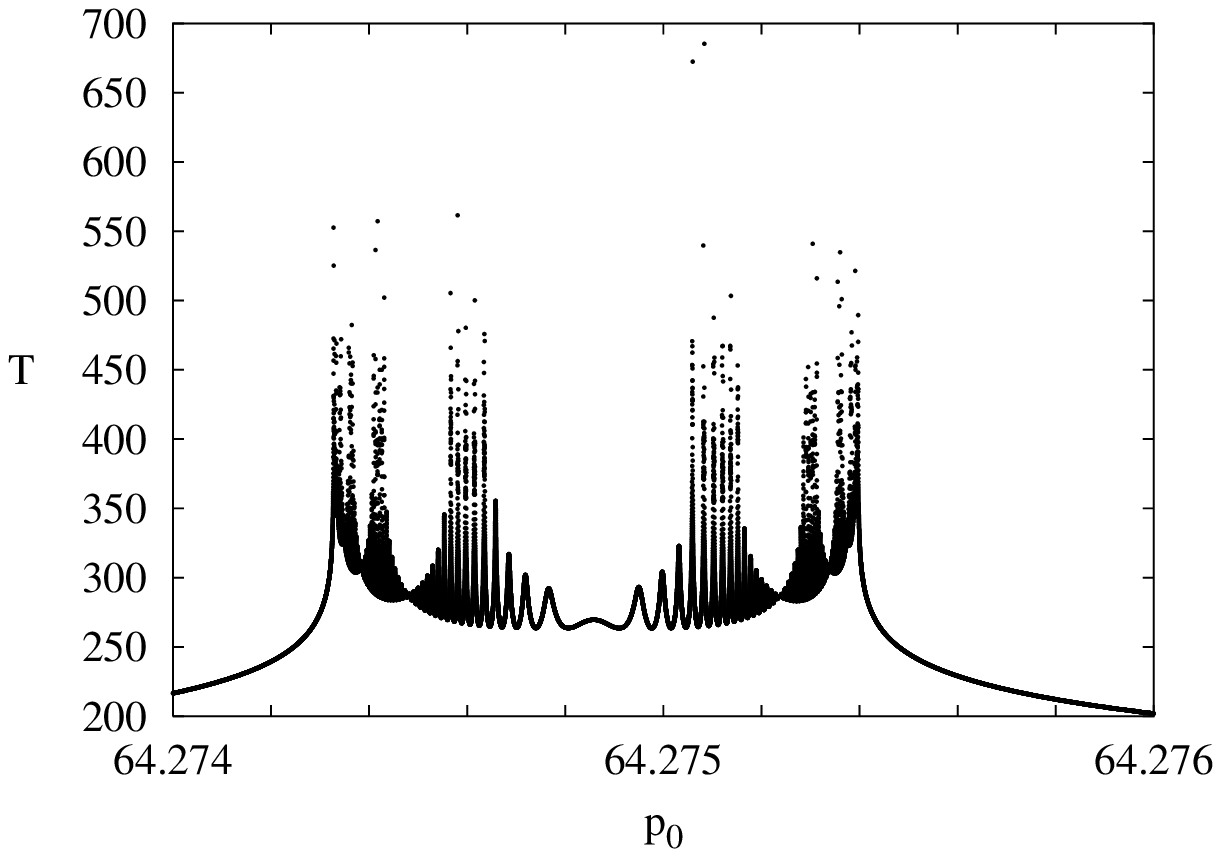}}
\caption{Atomic fractal in a quantized Fock field
with different resolutions ($\delta=0.4$ and $z(0)=0$).}\label{fig6}
\end{center}
\end{figure}
\begin{figure}[h!t]
\begin{center}
{\bf (a)} \parbox[t]{0.6\textwidth}{\rule{0pt}{0pt}\\
\includegraphics[width=0.6\textwidth]{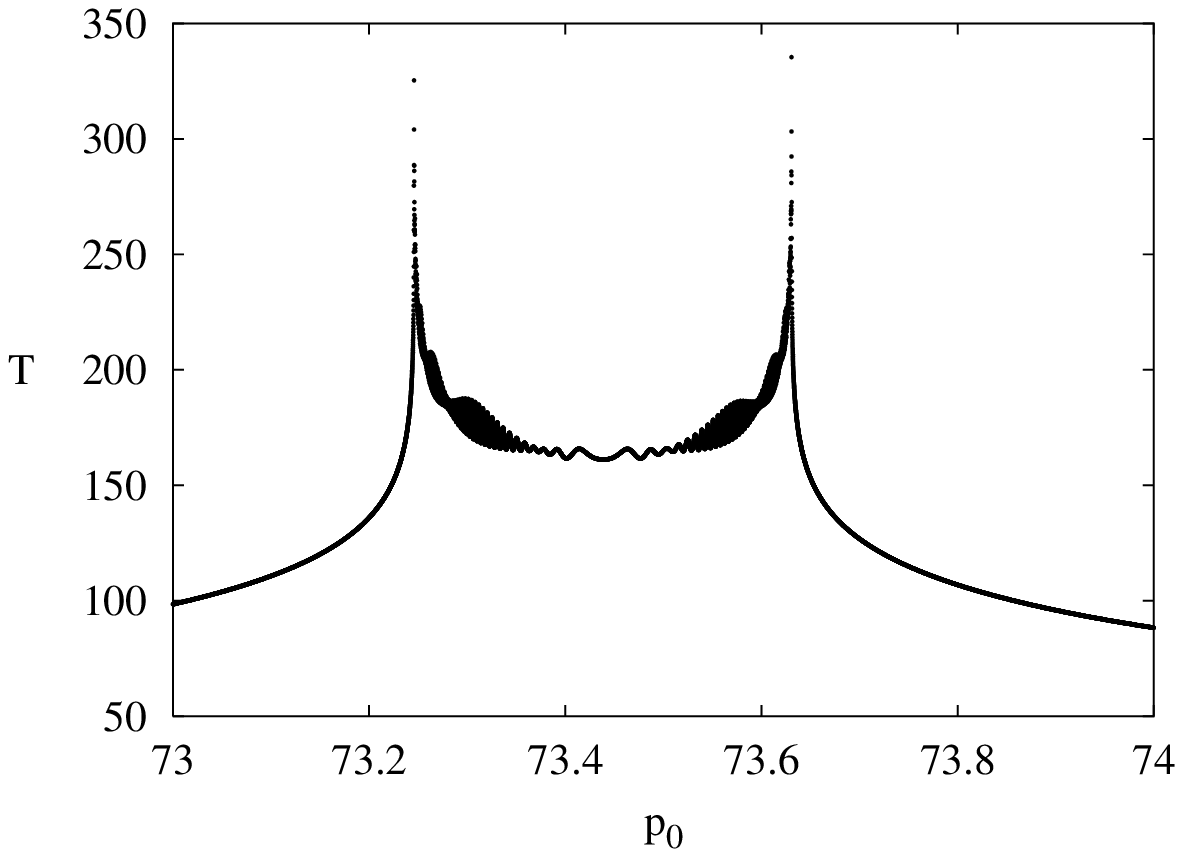}}\\
{\bf (b)} \parbox[t]{0.6\textwidth}{\rule{0pt}{0pt}\\
\includegraphics[width=0.6\textwidth]{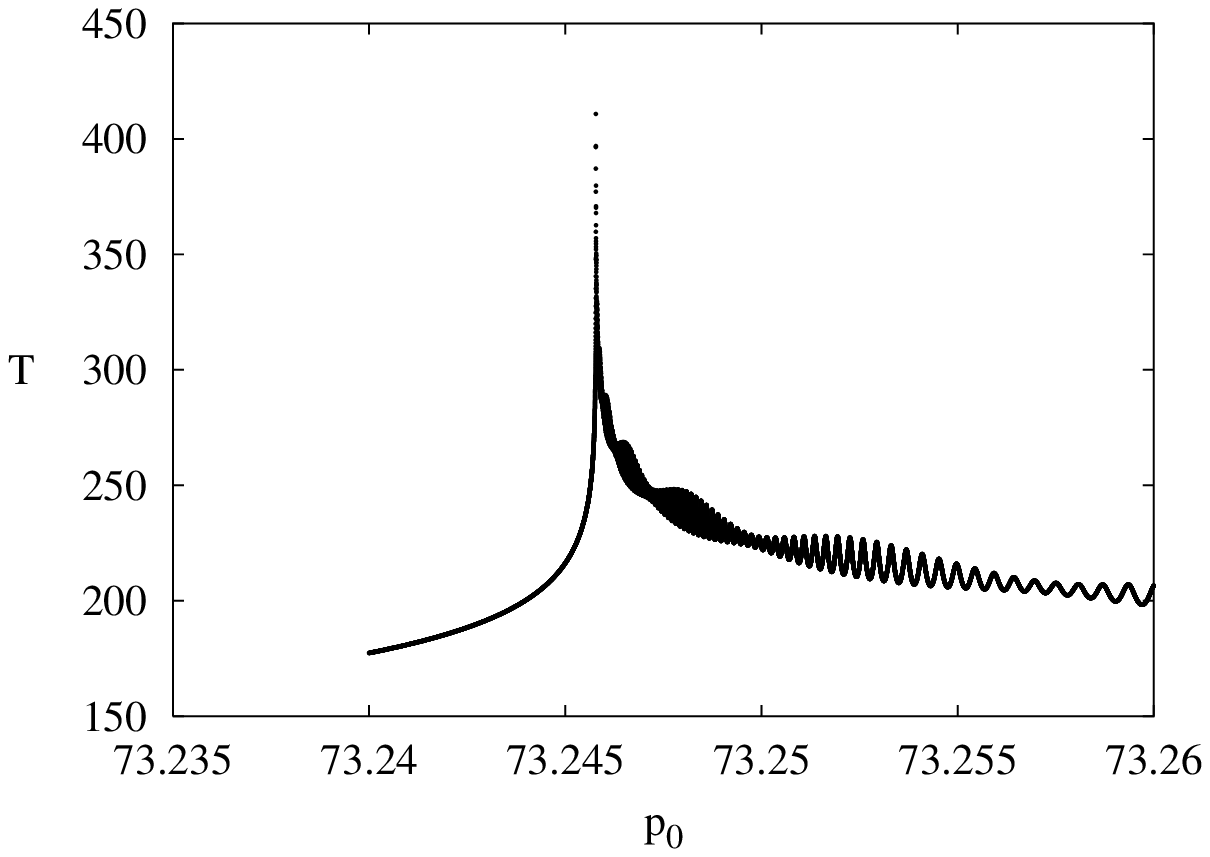}}
\caption{Example of one of the non-fractal substructures in the atomic fractal
shown in Fig.~\ref{fig6}a with successive magnifications.}\label{fig7}
\end{center}
\end{figure}
At exact resonance ($\delta=0$) with $u_{n-1}(\tau)=u_{n}(\tau)=0$,
the optical potential $U=(\sqrt{\bar n}\,u_{n-1}+\sqrt{\bar n+1}\,u_n)\cos x-
\delta(z_{n-1}+z_{n})/2$ is equal to zero, and
the analytical expression for the dependence in
question can be easily found to be the following:
$T(\delta=0)=3\pi/2\alpha p_0$ if $p_0>0$ and
$T(\delta=0)=\pi/2\alpha p_0$ if $p_0<0$.
Atoms simply fly through the cavity in one direction with their initial constant
velocity and are registered by one of the detectors.
Out of resonance ($\delta\ne 0$), the atomic motion has been numerically found
in the preceding section to be chaotic with
positive values of the maximal Lyapunov exponent. Fig.~\ref{fig6} shows
the function $T(p_0)$ with the normalized detuning $\delta=0.4$, the
recoil frequency $\alpha=10^{-3}$, and the average initial number of
cavity photons $\bar n=10$.
The exit-time function demonstrates an intermittency of smooth
curves and complicated structures that cannot be resolved in principle, no
matter how large the magnification factor.
Fig.~\ref{fig6}b shows magnification of the function for the small interval
$64.1\leqslant p_0\leqslant 64.6$. Further magnification in the range
$64.2743\leqslant p_0\leqslant 64.2754$ shown in Fig.~\ref{fig6}c  reveals
a beautiful self-similar structure.
Some structures in Fig.~\ref{fig6}a that look like fractal are not, in fact,
unresolvable and self-similar. Magnification of the structure in the range
$73.2\leqslant p_0\leqslant 73.8$ (Fig.~\ref{fig7}a), shown in
Fig.~\ref{fig7}b, demonstrates quite a smooth function without unresolvable
substructures and with only two singular points on the borders of the
respective momentum interval.
Beating, visible in all the structures
of the atomic fractal in Figs.~\ref{fig6},~\ref{fig7}, should be attributed
to the structure of the Hamilton-Schr\"odinger equations (\ref{24}) which
describe two atom-field oscillators with slightly different frequencies.

\begin{figure}[h!t]
\begin{center}
\includegraphics[width=0.8\textwidth]{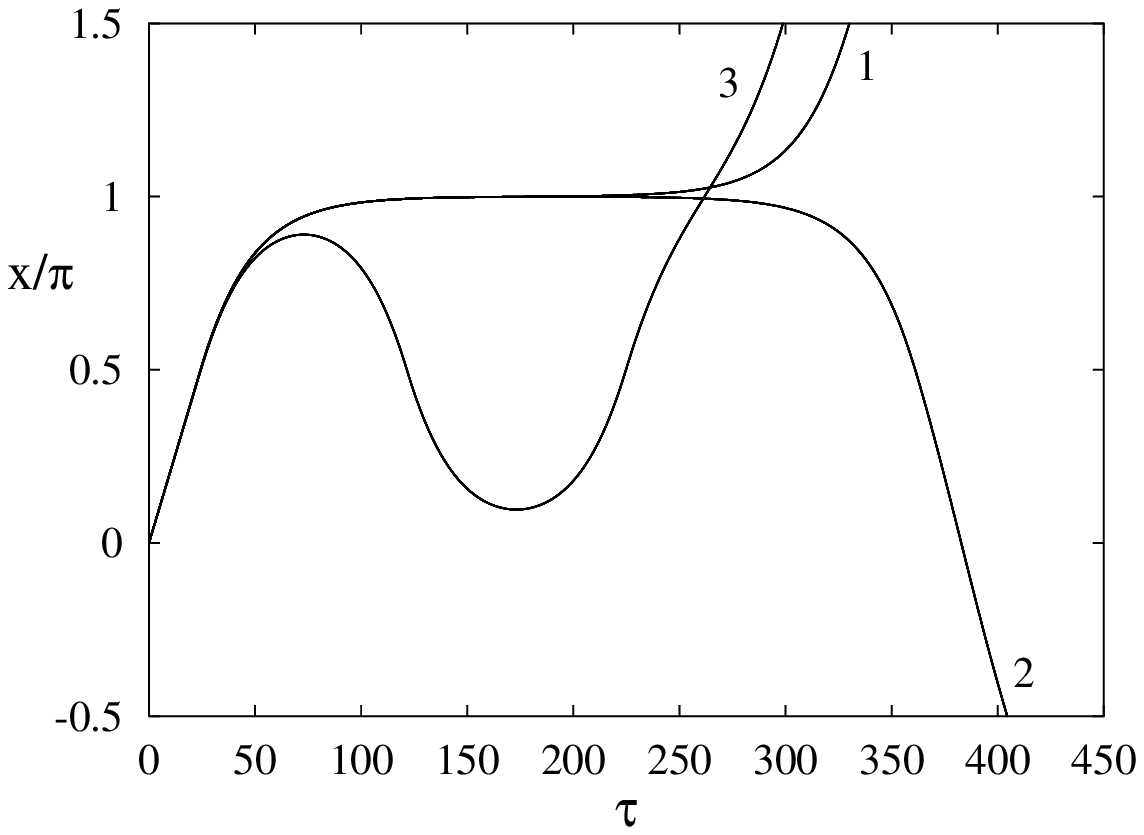}
\caption{Sample atomic trajectories transversing the central node in the
scheme Fig.~\ref{fig4}b $m$ times ($m=1,\,2,\,3$). The trajectory with
number 1 is close to a separatrix-like $1S$-trajectory.}\label{fig8}
\end{center}
\end{figure}
The exit time
$T$, corresponding to both smooth and unresolved $p_0$ intervals, increases
with increasing the magnification factor. It follows that there exist atoms
never reaching the detectors in spite of the fact that they have no obvious
energy restrictions to leave the cavity. Tiny interplay between chaotic external
and internal dynamics prevents these atoms from leaving the cavity. The similar
phenomenon in Hamiltonian systems is known as {\it dynamical trapping}
\cite{Z02}. Different
kinds of atomic trajectories, which are computed with the system (\ref{24}),
are shown in Fig.~\ref{fig8}. A trajectory with the number $m$ transverses
the central node of the standing-wave, before being detected, $m$ times and is
called $m$-th trajectory. There are also special separatrix-like
$mS$-trajectories following which atoms in infinite time reach the
stationary points $x_s=\pm\pi n$ ($n=0,\, 1,\, 2,\,\dots $),
$p_s=0$, transversing $m$ times the central node. These points are
the anti-nodes of the standing wave where the force acting on atoms is zero.
A detuned atom can
asymptotically reach one of the stationary points after transversing the
central node $m$ times.
The trajectory with number 1, showing in Fig.~\ref{fig8}, is close to a
separatrix-like $1S$-trajectory.
The smooth $p_0$ intervals in the first-order
structure in Fig.~\ref{fig6}a correspond to atoms transversing
once the central node and reaching the right detector. The unresolved singular
points in the first-order structure with $T=\infty$ at the border between the
smooth and unresolved $p_0$ intervals are generated by the
$1S$-trajectories. Analogously, the smooth and unresolved $p_0$ intervals
in the second-order structure in Fig.~\ref{fig6}b correspond to
the 2-nd order and the other trajectories, respectively, with singular points between them
corresponding to the $2S$-trajectories and so on.

There are two different mechanisms of generation of infinite exit times,
namely,
dynamical trapping with infinite oscillations ($m=\infty$) in a cavity and the
separatrix-like motion ($m\ne\infty$). The set of all initial momenta generating
the separatrix-like trajectories is a countable fractal. Each point in the set
can be specified as a vector in a Hilbert space with $m$ integer nonzero
components. One is able to prescribe to any unresolved interval of
$m$-th order structure a set with $m$ integers, where the first integer is a
number of a second-order structure to which trajectory under consideration
belongs in the first-order structure, the second integer is a number of a
third-order structure in the second-order structure mentioned above, and so on.
Such a number set is analogous to a directory tree address:
``$<$a subdirectory of the root directory$>$/$<$a subdirectory of the 2-nd
level$>$/$<$a subdirectory of the 3-rd level$>$/...''.

\section{Conclusion}

Atoms in cavities provide a microscopic nonlinear dynamical system with a
reach variety of qualitatively different dynamics that may be explored to
study the key problems of modern quantum physics and nonlinear science
including the problem of the quantum-classical correspondence.
The up-to-date experimental state of art has reached the stage where the
quantum to classical transition and the borderland between them can now be
probed directly.
Increasing the
number of atoms or/and the average number of photons in a cavity mode, one
can force the atom-field system to operating in quantum, semiquantum, and
semiclassical regimes providing a link between micro-, meso-, and
. In this paper, we derived the Hamilton -- Schr\"odinger
equations of motion that describe the semiquantum Hamiltonian dynamics of a
two-level atom with recoil strongly coupled to a single-mode standing-wave
quantized field in an ideal cavity. It has been shown that dynamical chaos
and fractals may arise in a wide range of the control parameters of the
system. To estimate their magnitudes we use the parameters of the real
experiments with single atoms in Fabry -- Perot microcavities in the
strong-coupling regime \cite{HL}, for which the amplitude of the atom-field
coupling strength may reach $\Omega_0=2\pi\cdot(10^7\div 10^8)$ Hz exceeding
the decay rates of the cavity mode and the atomic dipole. With the above
mentioned values of $\Omega_0$ and $k_f\simeq 2\pi\cdot 10^6$ m${}^{-1}$, one
can estimate the normalized recoil frequency $\alpha$ to be in range
$10^{-5}$--$10^{-2}$ depending on the magnitude of the atomic mass.

\section*{Acknowledgments}
This work was supported by the Russian Foundation for Basic Research under
Grant Nos. 02--02--17796, 02--02--06840, and 02--02--06841. One of the authors
(S. P.) thanks the Organizing Committee of the Workshop for supporting
his visit to Carry le Rouet.

\end{document}